\documentclass[twocolumn]{aastex631}
\usepackage{amsmath}
\usepackage{color, colortbl}

\newcommand{\mbh}{${\cal M}_{\mathrm{BH}}$}

\newcommand{\mstar}{${\cal M}_{\mathrm{*}}$}
\newcommand{\mdm}{${\cal M}_{\mathrm{DM}}$}

\newcommand{\kps}{${\rm km\,s}^{-1}$}

\definecolor{Gray}{gray}{0.9}
\shorttitle{Schwarzschild Modeling of NGC 4486B}
\shortauthors{B. Tahmasebzadeh et al.}
\graphicspath{{./}{figures/}}

\begin{document}

\title{A JWST View of the Overmassive Black Hole in NGC 4486B}

\author[0000-0002-1584-2281]{Behzad Tahmasebzadeh}
\affiliation{Department of Astronomy, 
University of Michigan, 
1085 S. University Ave., 
Ann Arbor, MI 48109, USA}

\author[0000-0003-3009-4928]{Matthew A.\ Taylor}
\affiliation{University of Calgary,
2500 University Drive NW,
Calgary, Alberta, T2N 1N4, Canada}

\author[0000-0002-6257-2341]{Monica Valluri}
\affiliation{Department of Astronomy, 
University of Michigan, 1085 S. University Ave., 
Ann Arbor, MI 48109, USA}

\author[0009-0006-9760-9315]{Haruka Yoshino}
\affiliation{University of Calgary,
2500 University Drive NW,
Calgary, Alberta, T2N 1N4, Canada}

\author[0000-0002-5038-9267]{Eugene Vasiliev}
\affiliation{University of Surrey, Guildford, GU2 7XH, UK}

\author[0000-0003-4867-0022]{Michael J.\ Drinkwater}
\affiliation{School of Mathematics and Physics, 
University of Queensland, 
Brisbane, QLD 4072, Australia}

\author[0009-0006-7485-7463]{Solveig Thompson}
\affiliation{University of Calgary,
2500 University Drive NW,
Calgary, Alberta, T2N 1N4, Canada}

\author[0000-0002-8532-4025]{Kristen Dage}
\affiliation{International Centre for Radio Astronomy Research -- Curtin University, GPO Box U1987, Perth, WA 6845, Australia}

\author[0000-0003-1184-8114]{Patrick C\^ot\'e}
\affiliation{National Research Council of Canada, 
Herzberg Astronomy and Astrophysics Program, 
5071 West Saanich Road, 
Victoria, BC, V9E 2E7, Canada}

\author[0000-0002-8224-1128]{Laura Ferrarese}
\affiliation{National Research Council of Canada, 
Herzberg Astronomy and Astrophysics Program, 
5071 West Saanich Road, 
Victoria, BC, V9E 2E7, Canada}

\author[0000-0002-0647-718X]{Tatsuya Akiba}
\affiliation{JILA and Department of Astrophysical and Planetary Sciences, CU Boulder, Boulder, CO 80309, USA}

\author[0000-0003-4703-7276]{Vivienne Baldassare}
\affiliation{Department of Physics and Astronomy, Washington State University, Pullman, WA 99163, USA}

\author[0000-0002-2816-5398]{Misty C.\ Bentz}
\affiliation{Department of Physics and Astronomy, Georgia State University, Atlanta, GA 30303, USA}

\author[0000-0002-5213-3548]{John P.\ Blakeslee}
\affiliation{NSF's National Optical-Infrared Astronomy Research Laboratory, 950 North Cherry Avenue, Tucson, AZ 85719, USA}

\author[0000-0002-1959-6946]{Holger Baumgardt}
\affiliation{School of Mathematics and Physics, 
The University of Queensland, 
St. Lucia, QLD 4072, Australia}

\author[0000-0001-6333-599X]{Youkyung Ko}
\affiliation{Korea Astronomy and Space Science Institute, 776 Daedeok-daero, Yuseong-Gu, Daejeon 34055, Republic of Korea}

\author[0000-0002-4718-3428]{Chengze Liu}
\affiliation{Department of Astronomy, School of Physics and Astronomy, and Shanghai Key Laboratory for Particle Physics and Cosmology, Shanghai Jiao Tong University, Shanghai 200240, People's Republic of China}

\author[0000-0002-1119-5769]{Ann-Marie Madigan}
\affiliation{JILA and Department of Astrophysical and Planetary Sciences, CU Boulder, Boulder, CO 80309, USA}

\author[0000-0002-2073-2781]{Eric W. Peng}
\affiliation{NSF's National Optical-Infrared Astronomy Research Laboratory, 
950 North Cherry Avenue, 
Tucson, AZ 85719, USA}

\author[0000-0002-0363-4266]{Joel Roediger}
\affiliation{National Research Council of Canada, 
Herzberg Astronomy and Astrophysics Program, 
5071 West Saanich Road, 
Victoria, BC, V9E 2E7, Canada}

\author[0000-0002-3382-9021]{Kaixiang Wang}
\affiliation{Department of Astronomy, 
Peking University, 
Beijing 100871, People's Republic of China}
\affiliation{Kavli Institute for Astronomy and Astrophysics, 
Peking University, 
Beijing 100871, People's Republic of China}

\author[0000-0003-1428-5775]{Tyrone E.\ Woods}
\affiliation{Department of Physics and Astronomy,
University of Manitoba,
30A Sifton Road,
Winnipeg, Manitoba, R3T 2N2 Canada}

\begin{abstract}
We present a new stellar dynamical measurement of the supermassive black hole in the compact elliptical galaxy NGC 4486B, based on integral field spectroscopy with JWST/NIRSpec. The two-dimensional kinematic maps reveal a resolved double nucleus and a velocity dispersion peak offset from the photometric center. Utilizing two independent methods—Schwarzschild orbit-superposition and Jeans Anisotropic Modeling—we tightly constrain the black hole mass by fitting the full line-of-sight velocity distribution.  
Our axisymmetric Schwarzschild models yield a best-fit black hole mass of \mbh$ = 3.6^{+0.7}_{-0.7} \times 10^8 \, M{\odot}$, slightly lower but significantly more precise than previous estimates. However, since our models do not account for the non-equilibrium nature of the double nucleus, this value may represent a lower limit. Across all tested dynamical models, the inferred \mbh$/M_*$ ratio ranges from $\sim$ 4–13\%, providing robust evidence for an overmassive SMBH in NGC 4486B.
Combined with the galaxy's location deep within the Virgo Cluster, our results support the interpretation that NGC 4486B is the tidally stripped remnant core of a formerly massive galaxy. As the JWST/NIRSpec field of view is insufficient to constrain the dark matter halo, we incorporate archival ground-based long-slit kinematics extending to $5\arcsec$. While this provides some leverage on the dark matter content, the constraints remain relatively weak. We place only an upper limit on the dark matter fraction, with \mdm/\mstar $\lesssim 0.5$ within 1 kpc—well beyond the effective radius. The inferred black hole mass remains unchanged with or without a dark matter halo.


\end{abstract}


\section{Introduction}
Compact elliptical (cE) galaxies are a class of compact stellar systems (CSSs) with radii ranging from a few hundred to a few thousand parsecs. Despite their small size, they possess substantial masses between $10^{8}$ and $10^{10} \, M_{\odot}$. The stars in these galaxies are predominantly old, suggesting that the majority of their star formation occurred long ago \citep{Chilingarian.2009S}.

Debate continues over how cEs form. Those cEs near larger host galaxies typically display redder colors, smaller sizes, and older stellar populations compared to isolated ones. These variations in observed properties suggest that cEs may originate through multiple mechanisms. Some might gradually accumulate stellar mass in isolation \citep{Zolotov.2015}, while others could result from tidal stripping by a more massive neighboring galaxy \citep{Bekki.2001, Deeley.2023}, or the mean cluster tidal field \citep{Valluri.1993, Gnedin.2003}. Recent high-resolution simulations further support the role of environmental processes in forming compact, metal-rich systems: \citet{Du.2019} demonstrate that ram-pressure confinement during pericentric passages can trigger centrally concentrated starbursts, enhancing compactness and metallicity, while \citet{Bian.2025}, using the TNG50 simulation \citep{Nelson.2019}, find that the majority of metal-rich compact stellar systems form via similar bursty star formation episodes driven by ram pressure and tidal forces, rather than purely through tidal stripping.

The presence of dark matter (DM) in cEs remains contentious. While some studies report substantial amounts of DM in these cEs, others find no evidence of DM, highlighting the possibility that cEs may arise through diverse formation mechanisms \citep{Yildirim.2017, Buote.2019}.

In the stripped-nucleus formation scenario, cEs are expected to retain the central supermassive black holes (SMBHs) of their progenitor galaxies. Observations within the Local Volume have identified candidate massive black holes in nearby cEs, including NGC 221 (M32) with estimated masses of \mbh$ \approx 3 \times 10^6 \, M_\odot$ \citep{Bender.1996, Verolme.2002}. A more massive candidate has been suggested in the Virgo cluster cE NGC 4486B, where \citet{Kormendy.1997} proposed a black hole (BH) mass of \mbh $\approx 6^{+3}_{-2} \times 10^8 \, M_\odot$ based on spherical isotropic Jeans modeling; however, the detection remained unconfirmed under their anisotropic model.  


Stellar dynamical modeling, particularly the orbit-superposition method \citep{sch_1979}, is a robust approach for constraining central BH masses in nearby galaxies when combined with data from integral field spectroscopic instruments.  Among the measured BHs in low mass galaxies, only \cite{Verolme.2002} used the orbit-superposition method, and the rest \citep{Kormendy.1997, Nguyen.2019, Seth.2010, Davis.2020} uses Jeans Anisotropic MGE (JAM) modeling \citep{Cappellari.2020}.

In this study, we investigate the presence of a BH in NGC 4486B using JWST/NIRSpec IFU data and the orbit-superposition method implemented in the FORSTAND code \citep{Vasiliev.2020A}, which is part of the AGAMA stellar dynamics toolbox \citep{agama}. To assess potential biases associated with different modeling approaches, we also perform JAM.  The FORSTAND code has previously been used to measure central BH masses in two disk galaxies \citep{Roberts.2021, Merrell.2023}. Furthermore, \citet{Behzad.2024} employed FORSTAND to rigorously assess the recovery of BH masses in mock compact stellar systems designed to resemble galaxies in the Virgo cluster, within the sensitivity range of JWST/NIRSpec IFU. This modeling framework also enabled the discovery of an overmassive BH in an ultra-compact dwarf galaxy (UCD736) in the Virgo cluster \citep{Taylor.2025}.

NGC 4486B is a cE galaxy located in the center of the Virgo Cluster, located $7.3^{\prime}$ ($\sim 35$ kpc) in projection from the massive elliptical M87, around which it orbits as a close satellite. It is characterized by a small effective radius of $2.33^{\prime\prime}$ ($\sim 0.19$ kpc) and an absolute magnitude of $M_V \sim –17.6$, with a total stellar mass of $\sim 6 \times 10^9 \, M_{\odot}$ \citep{Lauer.1996, Ferrarese.2006, Janz.2016}. These properties place NGC 4486B among the most extreme examples of tidally stripped systems, likely representing the dense remnant core of a once more massive progenitor.

HST observations have revealed that NGC 4486B hosts a double nucleus \citep{Lauer.1996}, analogous to the one observed in M31 \citep{Lauer.1993}. Photometrically, the brighter nucleus (P1) has a central surface brightness of $\mu_{I} = 12.89$ mag arcsec$^{-2}$, while the fainter nucleus (P2) has $\mu_{I} = 13.00$ mag arcsec$^{-2}$, with a minimum of $\mu_I = 13.06$ mag arcsec$^{-2}$ between them \citep{Lauer.1996}. One explanation for these double nuclei is the presence of an apsidally aligned, eccentric nuclear disk composed of stars orbiting within the near-Keplerian gravitational field of an SMBH \citep{Tremaine.1995, Peiris.2003, Sambhus.2002, Bacon.2001, Wernke.2021, Akiba.2021A}. In this model, the SMBH's gravitational influence dominates over that of the disk, and the brighter nucleus arises from stars accumulating near their orbital apocenters, while the fainter nucleus corresponds to stars near periapsis.

Alternative scenarios, such as an ongoing merger between the nucleus of NGC 4486B and a less luminous stellar system, are considered implausible. As argued by \citet{Lauer.1996}, the nuclear morphology is independent of color, suggesting it is not due to dust obscuration, and the dynamical friction timescale for such a merger would be short ($<10^8$\,yr). Furthermore, the dense environment around M87 is expected to inhibit mergers at small separations.

Dynamical models constructed by JAM and Schwarzschild orbit-superposition methods generally assume global symmetry and steady-state equilibrium, and therefore cannot reproduce the asymmetric kinematics and luminosity distribution associated with double nuclei. However, we adopted an axisymmetric modeling framework to constrain the BH mass in NGC 4486B and explore several plausible modeling scenarios that yield meaningful constraints, particularly in comparison to previous measurements. This approach follows common practice in the literature, where simplified models are often employed to extract key physical insights from systems that are intrinsically more complex \citep[e.g.,][]{Richstone1990, Kormendy.1997, Voggel.2022}.

The structure of this paper is as follows. In Section~\ref{sec:obs}, we describe the datasets used in our analysis. Section~\ref{sec:kin} details the extraction of the kinematic profiles. In Section~\ref{sec:model}, we present two dynamical modeling approaches applied to the kinematic data. The results are discussed in Section~\ref{sec:result}. Throughout this study, we adopt a distance of $16.3$ Mpc to NGC 4486B \citep{Blakeslee.2009}.

\section{OBSERVATIONS AND DATA PROCESSING}\label{sec:obs}
Observational data were collected from two space-based observatories: the James Webb Space Telescope (JWST) and the Hubble Space Telescope (HST). Spectroscopic observations were obtained with JWST during Cycle 1 (Program ID 2576; PI: M. Taylor) using the integral field unit mode of the Near-Infrared Spectrograph \citep[JWST/NIRSpec IFU;][]{Boker.2022, Jakobsen.2022}. 

To improve the accuracy of our light-profile deprojection and enable detailed photometric analysis, we incorporated imaging data from the HST. Observations were obtained with the Advanced Camera for Surveys Wide Field Channel (ACS/WFC), which offers a pixel scale of $\sim0.05\arcsec,\text{pix}^{-1}$—equivalent to $\lesssim 4$ pc at the assumed distance of NGC 4486B. This resolution provides a factor of $\gtrsim 2$ improvement over what can be achieved with the NIRSpec IFU white-light image alone. We use images taken in the F850LP filter (roughly corresponding to the SDSS $z'$ band) taken as part of the Advanced Camera for Surveys Virgo Cluster Survey program \citep{Cote.2004} with an integration time of 1210\,s. We use a point spread function (PSF) for the final combined image generated using the TinyTim software package \citep{Krist.2011}. For comparison, we also include the WFPC2 F555W image from \citet{Lauer.1996}, which provides slightly higher resolution in the central region and reveals the double nucleus structure more distinctly.

NGC 4486B was observed using the G235H grating and F170LP filter combination, covering the $1.66$–$3.17\,\mu$m wavelength range and providing a resolving power that increases from $R \approx 2200$ at $1.66\,\mu$m to $R \approx 4400$ at $3.17\,\mu$m. This resolution provides a velocity FWHM of $\Delta v_{\mathrm{los}} \approx 47 $\kps\ over the wavelength range $1.66\la\lambda/\mu m\la3.17$. Datacubes are approximately $3\arcsec$ on a side, with $0.1\arcsec$ spaxels, corresponding to a spatial scale of $\sim8$ pc at the adopted distance to the target. The final datacube was constructed from four exposures, totaling $\sim 700 $ seconds of integration time, using a four-point dither pattern with $0.025''$ sub-slice offsets to improve PSF sampling. Data were acquired using the NRSIRS2RAPID readout mode. Given the extended nature of NGC\,4486B, dedicated offset nods were taken of a nearby background field with the identical setup to enable accurate background subtraction.

 \begin{figure*}
	\centering	%
	\includegraphics[width=2\columnwidth]{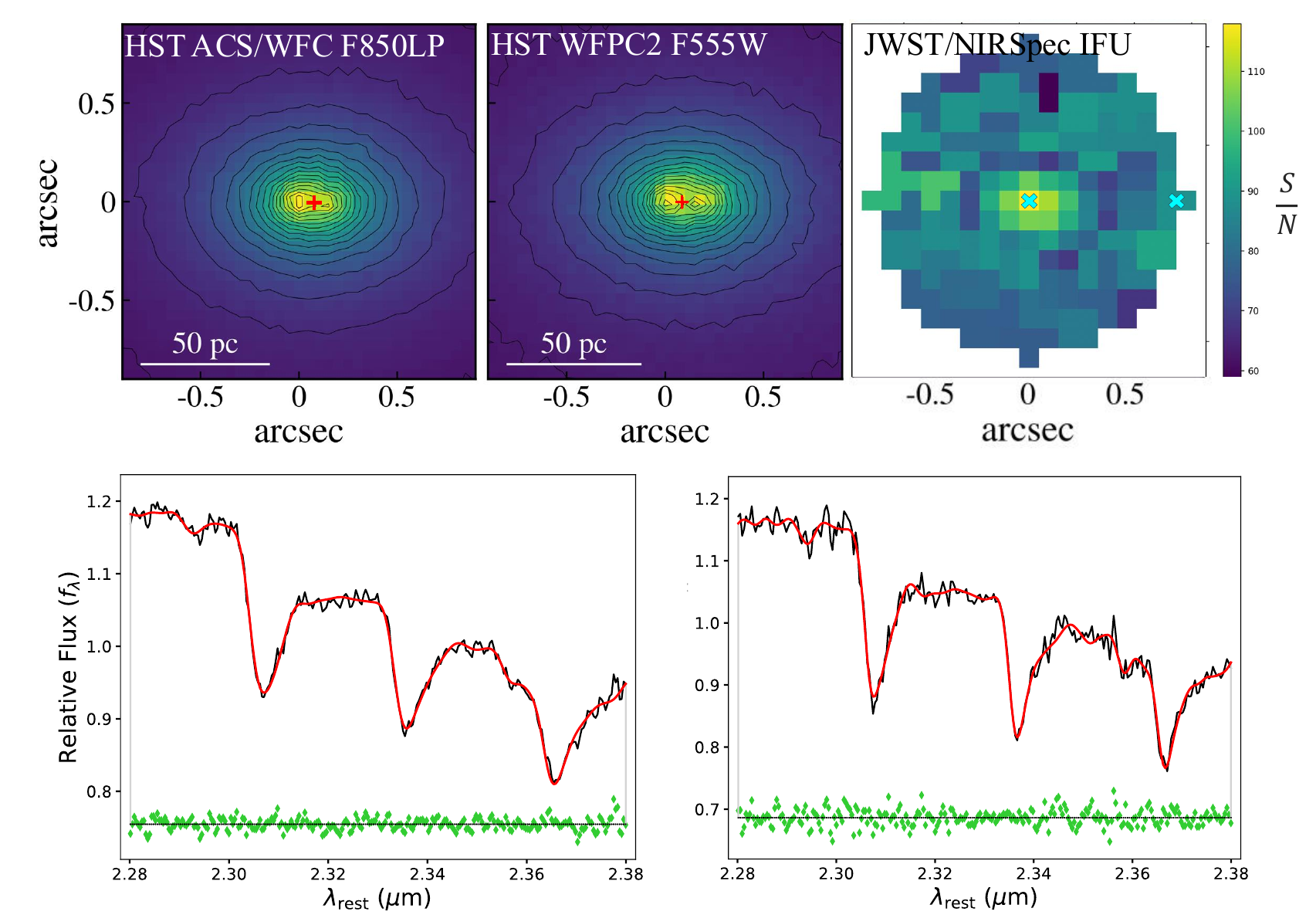}
	\caption{Top row: The left and middle panels show the deconvolved ACS/WFC F850LP and WFPC2 F555W images of NGC 4486B, respectively, with overlaid isophotal contours. The red plus marks the center of the outer isophote. Two distinct nuclei are visible in the WFPC2 F555W image, originally published in \citet{Lauer.1996}. The right panel shows the Voronoi-binned S/N map per \AA\ from the JWST/NIRSpec IFU data, revealing a central peak of $S/N \gtrsim 120$ that declines to $ S/N \gtrsim 48$ toward the field edges. The center of the IFU map corresponds to the photometric center (brightest light peak), which we adopt as the dynamical center for our analysis. The HST images are aligned such that the brightest light peak is placed at the center (0,0) of the coordinate system. All three panels cover the same spatial region. Bottom row: Example \texttt{pPXF} spectral fits for the central bin (left) and the outermost Voronoi bin (right), as marked by cyan crosses on the $S/N$ map. In each panel, the observed spectrum is shown in black, the best-fit model in red, and the residuals in green.}%
	\label{fig:ppxf}%
\end{figure*}


The datacubes were reduced using the STScI JWST/NIRSpec IFU pipeline (v1.17.1) with CRDS (Calibration Reference Data System) 
 context 1322. During Stage 1, the “snowball” flagging routine was enabled to detect major cosmic ray hits. This routine applied a $4\sigma$ rejection threshold and flagged events spanning at least 10 contiguous pixels, along with neighboring pixels affected by the impact. Following detector-level corrections, the individual dithered exposures proceeded through Stage 2, which included WCS (World Coordinate System) assignment, flat-fielding, and wavelength calibration. The resulting resampled 3D datacubes were then prepared for Stage 3, where they were combined into a final data product.

Before Stage 3 processing, additional cleaning steps were implemented to improve datacube quality and suppress residual spectral artifacts that persisted after default pipeline processing. These custom routines included flagging pixels in unreliable flat-field regions, identifying saturated pixels in individual frames, and correcting calibration-induced defects. A final check was performed to flag any pixels affected by failed open shutters in the NIRSpec micro-shutter array. Dedicated background frames were processed in an identical way to provide a master background frame that was subtracted during final datacube construction.

\section{Kinematic Measurements\label{sec:kin}}
To improve the spectral signal-to-noise ratio ($S/N$) and reduce the impact of anomalous IFU spaxels, we applied spatial binning using the Voronoi binning method described by \citet{Cappellari.2003}, targeting a minimum $S/N$ of approximately 80 per spectral pixel. The center of the cE was identified as the spaxel with the highest $S/N \approx 120$ per spectral resolution element, while the outermost Voronoi bin reached a minimum $S/N$ of $\sim48$. 

The top row of Fig. \ref{fig:ppxf} presents high-resolution HST images of the central region of NGC 4486B: the ACS/WFC F850LP image on the left and the WFPC2 F555W image in the middle, both with overlaid isophotal contours. The galaxy’s isophotal center (i.e., the geometric center from large-scale isophote fitting), marked by a red plus, lies between the two nuclear light peaks. The images are centered on the brightest peak, which corresponds to the center of the IFU field of view. The upper-right panel shows the Voronoi binned $S/N$ map of JWST/NIRSpec IFU data. All panels share the same spatial coverage ($2''\times2''$, corresponding to 158 pc $\times$ 158 pc). The full photometric image extends to a radius of 1 kpc, with an effective radius of 182 pc ($2.33''$), well beyond the coverage of the IFU data. The expected BH sphere of influence ($R_{\mathrm{infl}}$ in NGC 4486B is estimated to be $20-35$ pc, and is well resolved given the spatial resolution of the JWST/NIRSpec IFU data. 



The spatially binned spectra are used to extract the stellar kinematics from the integrated light, specifically the line-of-sight velocity distribution (LOSVD), using the penalized pixel-fitting software \citep[\texttt{pPXF};][]{Cappellari.2023}. Physically, we aim to recover the line-of-sight radial velocity ($v_{\mathrm{los}}$) and the line-of-sight velocity dispersion ($\sigma_{\mathrm{los}}$). \texttt{pPXF} parametrizes the LOSVD using a Gauss-Hermite (GH) series, characterized by the best-fitting Gaussian parameters: the mean velocity $v_{0}$, the width $\sigma_{0}$, and the higher-order moments $h_{3}$ (related to skewness) and $h_{4}$ (related to kurtosis) \citep[e.g.,][]{Marel.1993, Gerhard.1993}. Although $\sigma_{0}$ represents the width of the Gaussian base and does not correspond exactly to the physical second moment of the full LOSVD, for simplicity, we refer to $\sigma_{0}$ as the ``velocity dispersion" throughout the remainder of this paper. To generate model spectra, the software convolves a stellar template library with the parametrized LOSVD, as well as the line-spread function of the NIRSpec IFU in our adopted setup to match the spectral resolution of the JWST/NIRSpec IFU data.


We employed a library of $\sim600$ high-resolution synthetic spectra from the PHOENIX database \citep{Husser.2013}, covering a broad range of stellar parameters:
$3800 \leq T_{\rm eff} \leq 6000$\,K,
$-1.5 \leq {\rm [Fe/H]} \leq 1.0$\,dex,
$-0.2 \leq [\alpha/{\rm H}] \leq 2.0$\,dex,
$0.5 \leq \log g \leq 2.5$. 
Before fitting, both the observed and template spectra are truncated to the wavelength range $2.20$–$2.43 \, \mu{\rm m}$, focusing on the strong CO bandheads between $2.29$–$2.38 \, \mu{\rm m}$ for kinematic extraction (see Fig.~\ref{fig:ppxf}, bottom row).

Initial LOSVD parameters, including $v_{\mathrm{los}}$, are adopted based on values from the NASA/IPAC Extragalactic Database (NED)\footnote{\url{https://ned.ipac.caltech.edu}} and the velocity dispersion reported by \citet{Adrien.2015}. The optimal values of $v_{0}$, $\sigma_{0}$, and the GH moments $h_3$ and $h_4$, are then determined by minimizing the $\chi^2$ difference between the observed galaxy spectrum and the model. \texttt{pPXF} \, performs this optimization iteratively, applying a penalty function that minimizes the squared deviation of the LOSVD from a best-fitting Gaussian. This method allows the recovery of higher-order kinematic moments in regions with high $S/N$, while defaulting to a Gaussian profile in lower $S/N$ regions to maintain fit stability. We adopt a 4th-degree additive polynomial for the continuum, providing an adequate correction for accurately extracting stellar kinematics from the spectra. We use the kinematic maps in their unsymmetrized form, although symmetrization is a common preprocessing step in dynamical modeling \citep{Roberts.2021, Thater.2022, Quenneville.2022}.

Kinematic uncertainties were estimated using Monte Carlo simulations. For each bin, random Gaussian noise --- scaled to the 1$\sigma$ error of the corresponding spectrum --- was added, and the spectra were refitted using \texttt{pPXF}. The 1$\sigma$ error per bin was computed by summing the noise contributions of individual spaxels in quadrature, preserving the spectral resolution. The final uncertainties were determined as the standard deviation of the resulting kinematic measurements across the simulated realizations.  Since NIRSpec operations have begun, it has been found that additional systematic uncertainties can arise from low-frequency, wavelength-dependent sinusoidal artifacts in the extracted 1D IFU spectra, known as ``wiggles'', that can be corrected for in post-processing \citep[e.g.\ using the ``WICKED'' algorithm;][]{Dumont.2025}. Applying WICKED to the NGC\,4486B datacube we find that only two spaxels are affected, which after correction and limiting the spectral coverage to just the CO-bandhead region, results in derived kinematics differing by $\lesssim1\,{\rm km\,s^{-1}}$ than an uncorrected spectrum and are thus negligible in our overall uncertainty budget.

The bottom row of Fig.~\ref{fig:ppxf} presents example spectral fits to the CO bandheads in the wavelength range $2.28 \lesssim \lambda/\mu\mathrm{m} \lesssim 2.38$, demonstrating the fit quality for the central Voronoi bin (left) and the faintest outer bin (right).

\begin{figure*}
	\centering	%
	\includegraphics[width=2.\columnwidth]{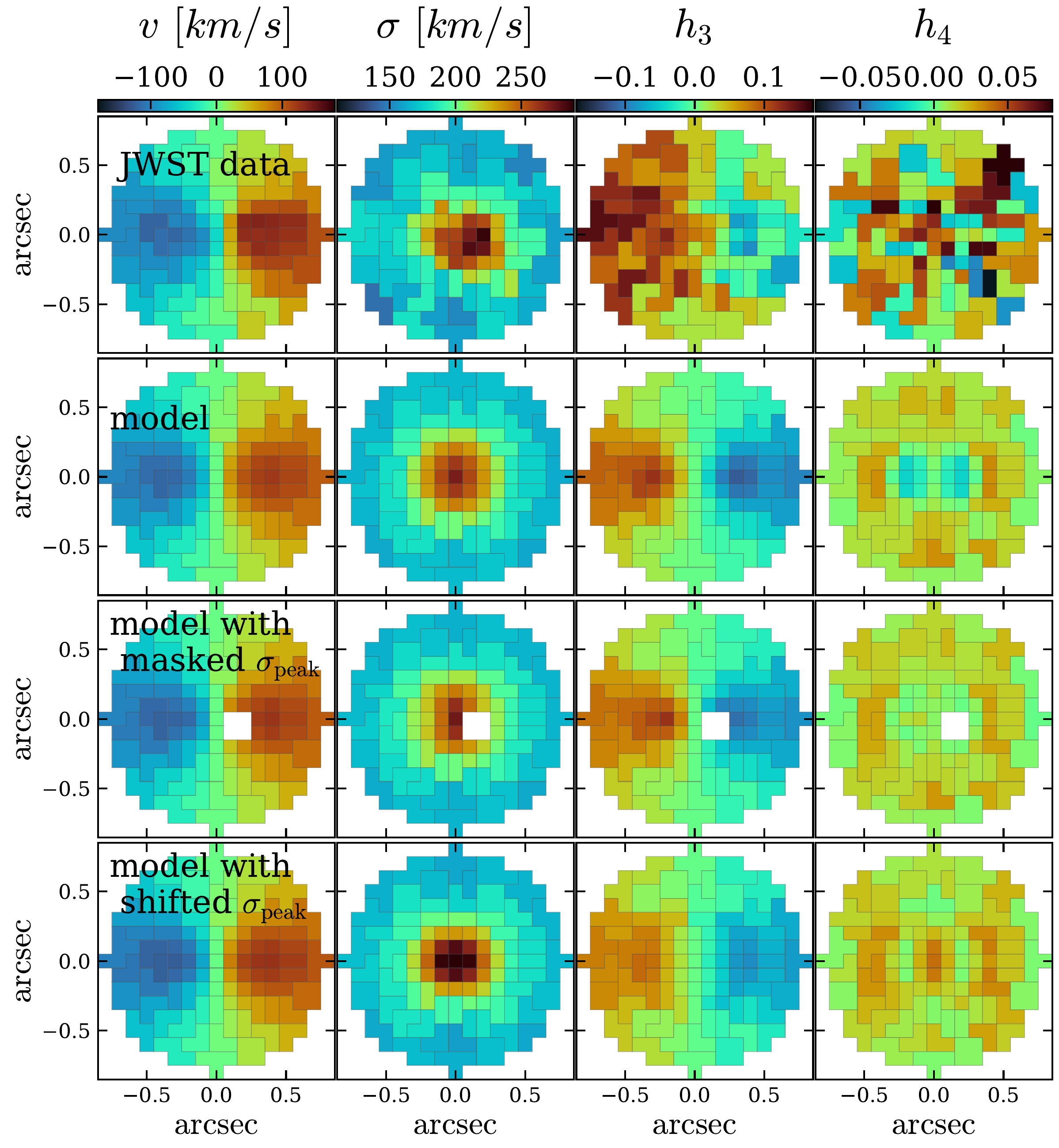}
	\caption{Kinematic maps represented by the GH coefficients ($v_0$, $\sigma_0$, $h_3$, $h_4$) are shown for NGC 4486B. The first row displays the JWST/NIRSpec IFU data, the second row shows the best-fitting Schwarzschild model based on the JWST dataset. The third row presents the best-fitting model for the modified data where the $\sigma$ peak is masked, and the fourth row shows the best-fitting model for data that the $\sigma$ peak values shifted to the center.}%
	\label{fig:kinematic}%
\end{figure*}

The first row of Fig.~\ref{fig:kinematic} presents the extracted 2D kinematic maps from the JWST/NIRSpec IFU data. The velocity dispersion $\sigma_{0}$ peaks at $287\, \mathrm{km \, s^{-1}}$, offset from the kinematic center, which coincides with the photometric center of the galaxy and exhibits a lower velocity dispersion $\sigma_{0}$ of $257\, \mathrm{km \, s^{-1}}$.

The spatial coverage of the JWST/NIRSpec IFU data is insufficient to constrain the DM component, which is crucial for understanding the formation pathway of this cE galaxy. To address this limitation, we incorporate additional one-dimensional kinematic data from \citet{Kormendy.1997}, obtained with the CFHT Subarcsecond Imaging Spectrograph (SIS; see their Table 2), using a $0.37\arcsec$-wide slit aligned with the galaxy’s double nucleus. However, discrepancies are observed between the JWST/NIRSpec IFU and CFHT/SIS kinematics in the inner region, particularly on the side where the velocity dispersion $\sigma_{0}$  peaks, where CFHT/SIS predicts higher velocity dispersion $\sigma_{0}$  and lower mean velocity $v_{0}$. The CFHT/SIS data show a central velocity dispersion $\sigma_{0}$  of $291\, \mathrm{km \, s^{-1}}$, but unlike the JWST/NIRSpec IFU data, they do not clearly resolve the offset in the velocity dispersion $\sigma_{0}$ peak. This discrepancy is likely due to the lower spatial resolution of the CFHT/SIS observations. From the F850LP image, the two nuclei of NGC 4486B are separated by 12\,pc, or $\sim 0.15 \arcsec$ \citep{Lauer.1996}, which is comparable to the SIS pixel size and significantly smaller than the observed seeing of FWHM $\sim 0.52\arcsec$.

Due to these limitations, we restrict the use of the CFHT/SIS kinematics to the outer regions, specifically beyond a radius of 0.9\arcsec, where they are used to constrain the extended mass distribution in our dynamical modeling.

\section{Dynamical Modeling}\label{sec:model}
In this section, we outline the methodology used for Schwarzschild and JAM dynamical modeling of the kinematic and imaging data. Instrumental resolution is explicitly accounted for when comparing model predictions to observational data. The JWST/NIRSpec IFU PSF is modeled as a sum of two circular Gaussians, based on stellar observations taken with the same instrument configuration (Program ID: 1364, PI: M.~Bentz). These Gaussians have dispersions of $\sigma = 0.07\arcsec$ and $\sigma = 0.24\arcsec$, contributing 85\% and 15\% of the total PSF, respectively \citep{Bentz_2025}. The HST F850LP PSF is modeled using three circular Gaussians with dispersions of $\sigma = 0.02\arcsec$, $0.07\arcsec$, and $0.3\arcsec$, contributing 82.3\%, 17.5\%, and 0.2\% of the total PSF, respectively.

For the CFHT/SIS kinematic data, we adopt a single Gaussian PSF with $\mathrm{FWHM} = 0.52\arcsec$ (corresponding to $\sigma \approx 0.22\arcsec$), based on the spectroscopic seeing reported in \citet{Kormendy.1997}. In our modeling, we simultaneously incorporate both the JWST/NIRSpec IFU and CFHT/SIS datasets to constrain the system's stellar kinematics.

As discussed previously, our dynamical modeling methods assume global symmetry and steady-state equilibrium.  Therefore, they cannot reproduce the asymmetric kinematics and luminosity distribution associated with the presence of double nuclei. To assess the impact of this limitation, we repeat our dynamical modeling using two modified JWST/NIRSpec IFU kinematic maps. In the first approach, we mask the spaxels at and around the peak of the $\sigma_0$ in all kinematic moments maps, thereby reducing the asymmetry in the kinematic map. This test allows us to explore the minimum BH mass consistent with the data and assess whether the BH can still be constrained at the $3\sigma$ level. In the second test, we manually shift the peak of $\sigma_0$ to the photometric center—modifying only the $\sigma_0$ map—to roughly symmetrize the velocity dispersion distribution. While this adjustment is not physically motivated, it serves as a limiting case to explore the maximum BH mass that could be inferred under the assumption of symmetry.

\subsection{Schwarzschild Modeling}
\subsubsection{HST Imaging and Stellar Density Profile}
To derive the three-dimensional luminosity density, we use the multi-Gaussian expansion (MGE) method \citep[]{Cappellari.2002} to model the surface brightness distribution from the photometric image. We adopt the slightly smoother ACS/WFC F850LP image, which is better suited for MGE fitting and axisymmetric dynamical modeling. The radial light profiles from both HST bands are identical, and the PSF of the kinematic map is larger than that of either HST image, minimizing the impact of this choice on the dynamical analysis. The MGE technique represents the observed surface brightness as a sum of two-dimensional elliptical Gaussians. These 2D MGE models are convolved with the HST instrumental PSF before being fitted to the observed image. To model NGC 4486B as an axisymmetric system, we restrict the fits to Gaussians without isophotal twists. The best-fit model is then deprojected into a 3D axisymmetric luminosity distribution, assuming an inclination angle ($\theta$) treated as a free parameter. The corresponding 3D mass density profile is obtained by scaling the deprojected luminosity density by a constant stellar mass-to-light ratio ($M/L$). The results of the MGE fits are presented in Table \ref{table:mge}.

\begin{table}
	\centering
	\footnotesize
	\begin{tabular}
    {l@{\hspace{20pt}}l@{\hspace{20pt}}l@{\hspace{20pt}}l@{\hspace{20pt}}l}
		\hline
		$ j $    &  $ L_{j}$ $ \mathrm{L_{\odot}pc^{-2}} $ &  $ \sigma^{\prime}_{j} $ $ (\mathrm{arcsec}) $    & $ q^{\prime}_{j} $   &  $ \Delta \psi_{j}^{\prime} $ $(^{\circ})$   \\
		\hline
        $1$  & 39629.74 & 0.14 & 0.64 & 0.0  \\
        $2$ & 37912.10  & 0.27 & 0.71 & 0.0  \\
        $3$ & 21126.61  & 0.51 & 0.73 & 0.0  \\
        $4$ & 9652.02  & 0.97 & 0.81 & 0.0  \\
        $5$ & 4134.22  & 2.01 & 0.88 & 0.0  \\
        $6$ & 943.24   & 4.50 & 0.90 & 0.0  \\			
		\hline
	\end{tabular}
	\\
	\parbox{\columnwidth}{\caption{Details of the MGE fit to the HST image of NGC 4486B in the axisymmetric limit, modeled with $j$ individual Gaussians, where $L_{j}$ is the central flux in units of $(\mathrm{L_{\odot}\,pc^{-2}})$, $\sigma^{\prime}_{j}$ is the projected width of each Gaussian (in arcseconds), and $q^{\prime}_{j}$ is the projected flattening. The twist angle $\Delta \psi^{\prime}_{j}$ is $0$ in this case.}
		\label{table:mge}}
\end{table}

\subsubsection{Gravitational Potential}
We construct the total gravitational potential by accounting for contributions from the axisymmetric stellar component, a central BH, and the DM halo. The stellar potential is derived from the inferred 3D stellar luminosity density distribution (as discussed in the previous subsection) by applying an assumed $M/L$ and solving the Poisson equation. The BH's potential is modeled using a Plummer profile with a fixed scale radius of $a = 10^{-4}$ kpc and mass \mbh. The DM component is modeled using a spherical NFW profile \citep{NFW.1996} with a scale radius $r_h$ and peak circular velocity $v_h$. 

Given the limited spatial coverage of our JWST kinematic data, there exists a degeneracy between $r_h$ and $v_{h}$, such that we can only meaningfully constrain the ratio $v_{h}/r_h$, which governs the inner slope of the potential. To break this degeneracy, we fix the scale radius to $r_h = 1$ kpc. This choice is motivated by expectations from $\Lambda$CDM simulations for galaxies with stellar masses $\sim10^9\,M_\odot$, where unstripped halos typically have $r_s \sim 3-10$ kpc \citep{Dutton.2014}. However, in tidally stripped systems such as compact ellipticals and nuclear remnants, the outer DM halo is expected to be largely removed by interactions with a more massive host, leaving behind a tightly bound stellar nucleus embedded in a truncated halo \citep{Bekki.2003, Pfeffer.2013}. In some cases, simulations suggest that nearly all of the DM may be stripped, resulting in a stellar nucleus that is effectively baryon-dominated \citep{Smith.2016}. 

As a result, our primary models incorporate four free parameters: $M/L$, \mbh, the inclination angle of the galaxy ($\theta$), and the DM peak circular velocity $v_h$.

\subsubsection{Construction of the orbit library}
In contrast to other Schwarzschild modeling codes, which typically assign orbit initial conditions on a regular grid based on integrals of motion, the FORSTAND code adopts a different approach by sampling orbital initial conditions randomly within the 6D phase space (see Section 2.4 in \citealt{Vasiliev.2020A} for details). Spatial positions are sampled uniformly from the intrinsic stellar density profile of the model. Corresponding velocities are then generated from a Gaussian distribution, where the dispersions depend on position and are determined by solving the anisotropic Jeans equation for the axisymmetric potential and density (e.g., \citealt{Cappellari.2008}). 
The anisotropy parameter $\beta_0$ was set to zero when generating the initial conditions. As noted in \citet{Behzad.2024}, varying $\beta_0$ in this context does not significantly affect the modeling results within our framework, as the final anisotropy is determined by fitting to the kinematic data.

For each set of model parameters, an orbit library is generated by integrating $N_{\rm orb} = 20,000$ orbits over 100 dynamical time within the defined potential. The LOSVD for each orbit is initially stored in three-dimensional data cubes, with two coordinates representing the image plane and the third corresponding to the velocity axis. These distributions are expressed using a basis of tensor-product B-splines of degree 2. They are then convolved with the spatial PSF and re-binned onto the grid of Voronoi apertures.

\subsubsection{Parameter grids and fitting procedure}
After constructing the orbit library, we find the orbital weights that (a) reproduce the 3D density discretized over a cylindrical grid of $20 \times 15$ in the $R$, $z$ plane, and (b) minimize the objective function $\mathcal{F} \equiv \mathcal{F}_{\text {kin }}+\mathcal{F}_{\text {reg}}$. The first term $\mathcal{F}_{\text {kin }}$ determines the goodness of fit to the kinematic constraints $(v_{0}$, $\sigma_{0}$, $h_3$, $h_4)$. The second term $\mathcal{F}_{\text {reg}}$ is the `regularization-term', which penalizes large differences between orbital weights $\boldsymbol{w}$ to avoid overfitting. We define $\mathcal{F}_{\text {reg }}=\lambda N_{\text {orb }}^{-1} \sum_{i=1}^{N_{\text {orb }}}\left(w_i / \bar{w}\right)^2$, where the mean orbit weight is $\bar{w} \equiv M_{\star} / N_{\text {orb }}$, and the regularization coefficient $\lambda$ controls the trade-off between accurately reproducing kinematic constraints (with a smaller $\lambda$) and the smoothness of the model (with a larger $\lambda$). We set $\lambda=15$, which is adequate to prevent overfitting and to ensure a reasonably smooth likelihood surface when using full kinematic information \citep{Behzad.2024}. Upon determining the optimal orbit weights, we compute the final $\chi^{2}$ to assess the fit with respect to the observed kinematic moments $(v_{0}, \sigma_{0}, h_3, h_4)$ and their corresponding measurement uncertainties. For additional details, see Section 5.1 in \citet{Roberts.2021}.


 \begin{figure*}
	\centering	%
	\includegraphics[width=2\columnwidth]{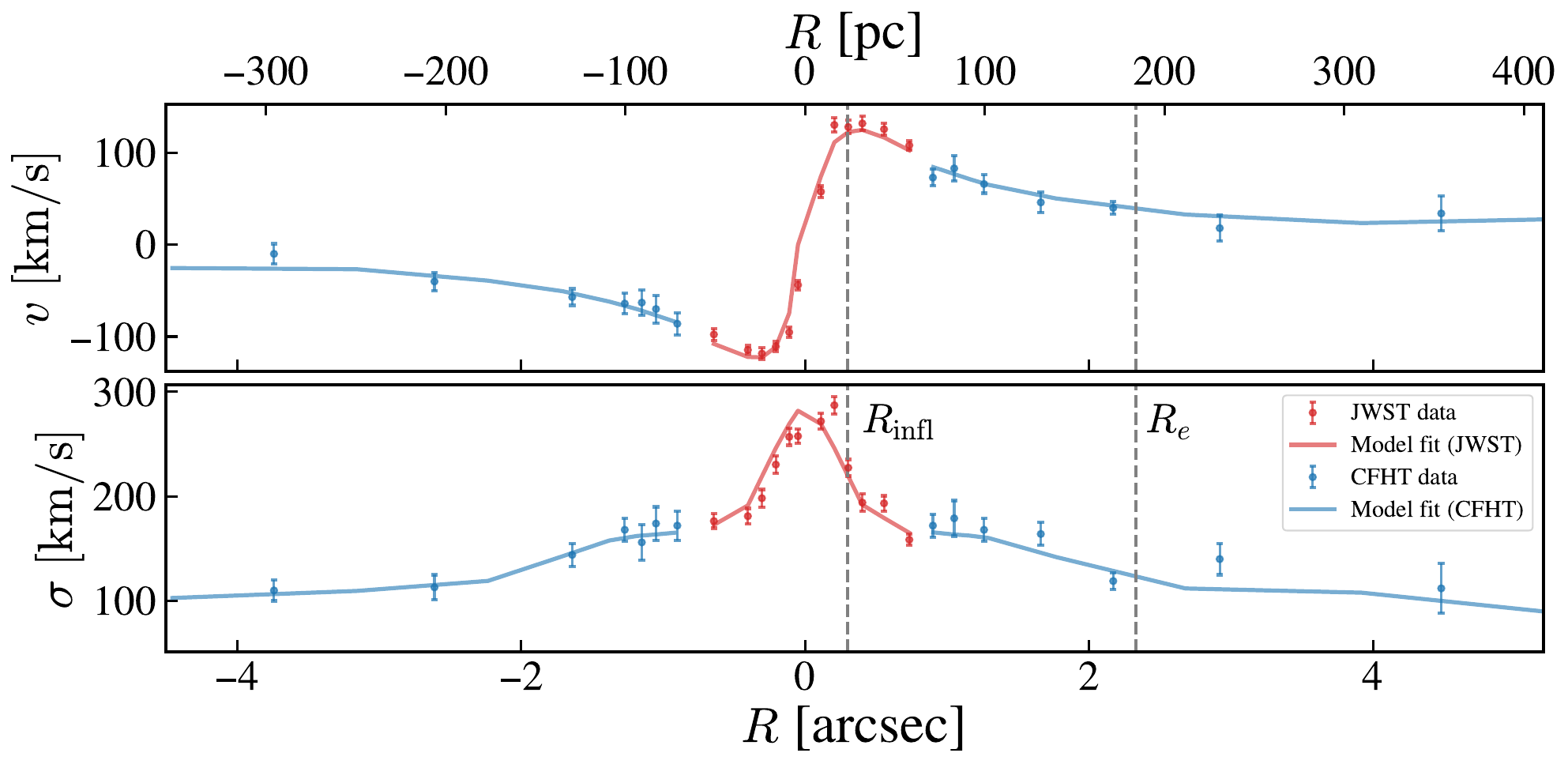}
	\caption{Kinematic profiles of $v_o$ and $\sigma_o$ along the kinematic major axis. Red points show the JWST/NIRSpec IFU data with their associated uncertainties, while the red line indicates the best-fit axisymmetric model. The blue points represent the CFHT/SIS data and uncertainties from \citet{Kormendy.1997}, and the blue line indicates the best-fit model. Vertical dashed lines mark the expected BH sphere of influence ($R_{\mathrm{infl}}$) and the galaxy’s effective radius ($R_e$), both are indicated relative to the center of our dynamical model, where the BH is assumed to reside at $R=0$. }%
	\label{fig:2set_kinem}%
\end{figure*}

To efficiently sample the multi-dimensional parameter space, we adopt a hypercube sampling \citep{McKay.1979} instead of performing model searches on a regular grid. This approach, similar to the method used by \cite{Quenneville.2022}, allows for more comprehensive coverage of the parameter space with fewer evaluations. We implement the sampling using the \texttt{LatinHypercube} routine from the Quasi-Monte Carlo submodule in the \texttt{SciPy} Python package. For each sampled combination of \mbh, $\theta$, and $v_h$, we construct a dedicated orbit library with initial $M/L$=1. Each orbit library is then reused to explore a range of $M/L$ by scaling the orbital velocities by $\sqrt{M/L}$. 

We begin with a Latin hypercube consisting of 5000 models, sampling the parameter space over the ranges $v_h \in [0, 200]$ $\mathrm{km \, s^{-1}}$, \mbh $\in [0.1, 15] \times 10^8\,M_{\odot}$, and $\theta \in [59^\circ, 90^\circ]$, corresponding to the inclination angles for which a valid deprojection of the luminosity density is possible. For each sampled parameter set, the $M/L$ is explored over the range $[0.5, 6]$ in multiplicative steps of 0.2. From this initial model grid, we compute the $1\sigma$, $2\sigma$, and $3\sigma$ confidence intervals for each parameter by marginalizing the $\chi^2$ distribution over all other parameters. To refine our measurements, we perform three additional iterations. First, we generate a new Latin hypercube of 20,000 models restricted to the $3\sigma$ confidence region. This is followed by a second iteration consisting of 10,000 models sampled within the $2\sigma$ region, and a final iteration of 10,000 models within the $1\sigma$ region of the existing sample. For the final iteration, we refine the $M/L$ grid by adopting multiplicative steps of $0.1$, which are half the size of those used in earlier iterations, in order to enhance the precision and robustness of the derived parameter constraints. In total, we compute $\sim \, 40000$ models.

We report the median values and corresponding uncertainties at the 68\%, 95\%, and 99.7\% confidence levels, derived from the marginalized posterior distributions of each parameter. These distributions are obtained by applying kernel density estimation to the full model grid weighted by $\exp(-\chi^2/2)$.

\begin{figure*}
	\centering	%
	\includegraphics[width=1\columnwidth]{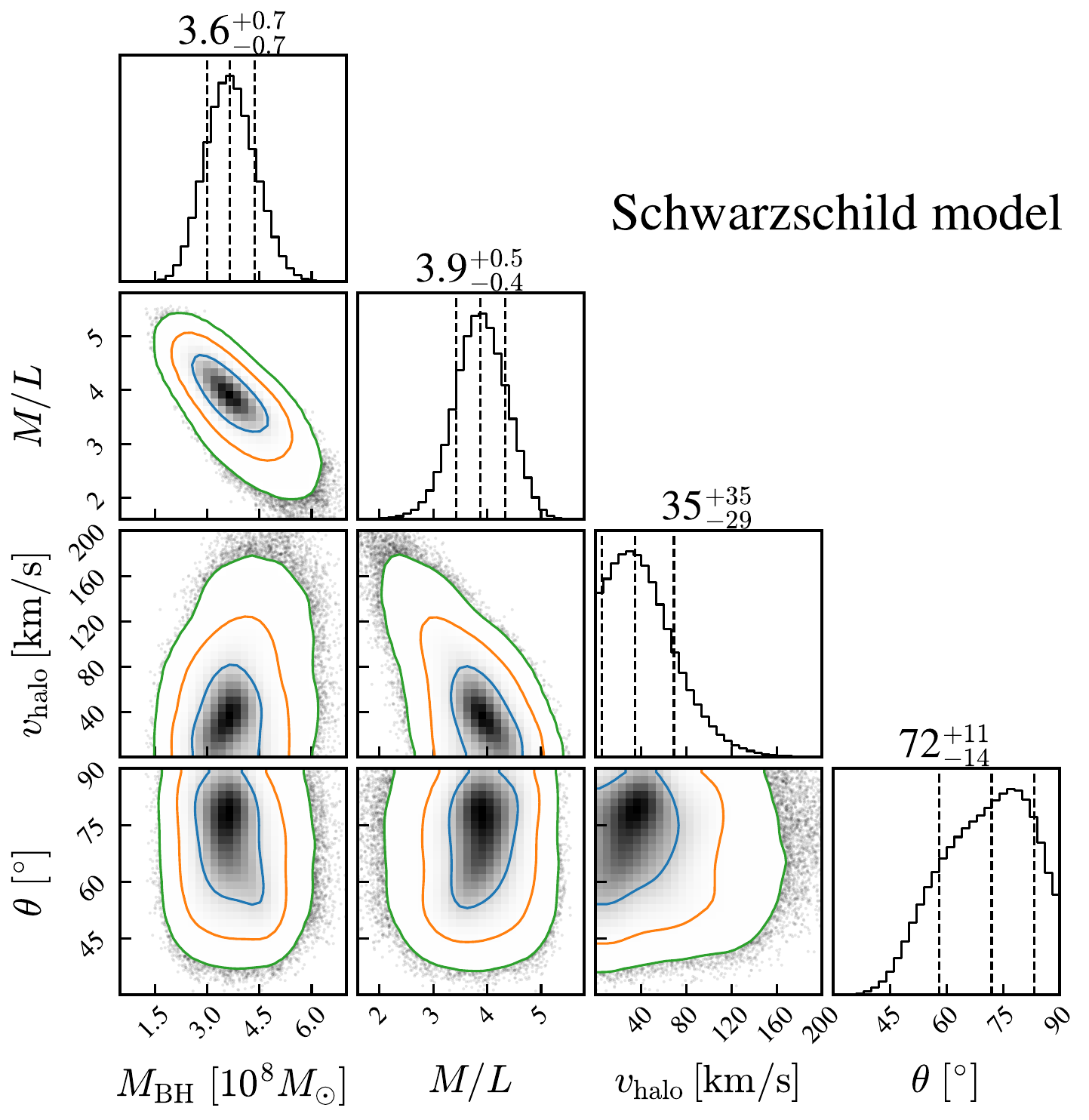}
    \includegraphics[width=1\columnwidth]{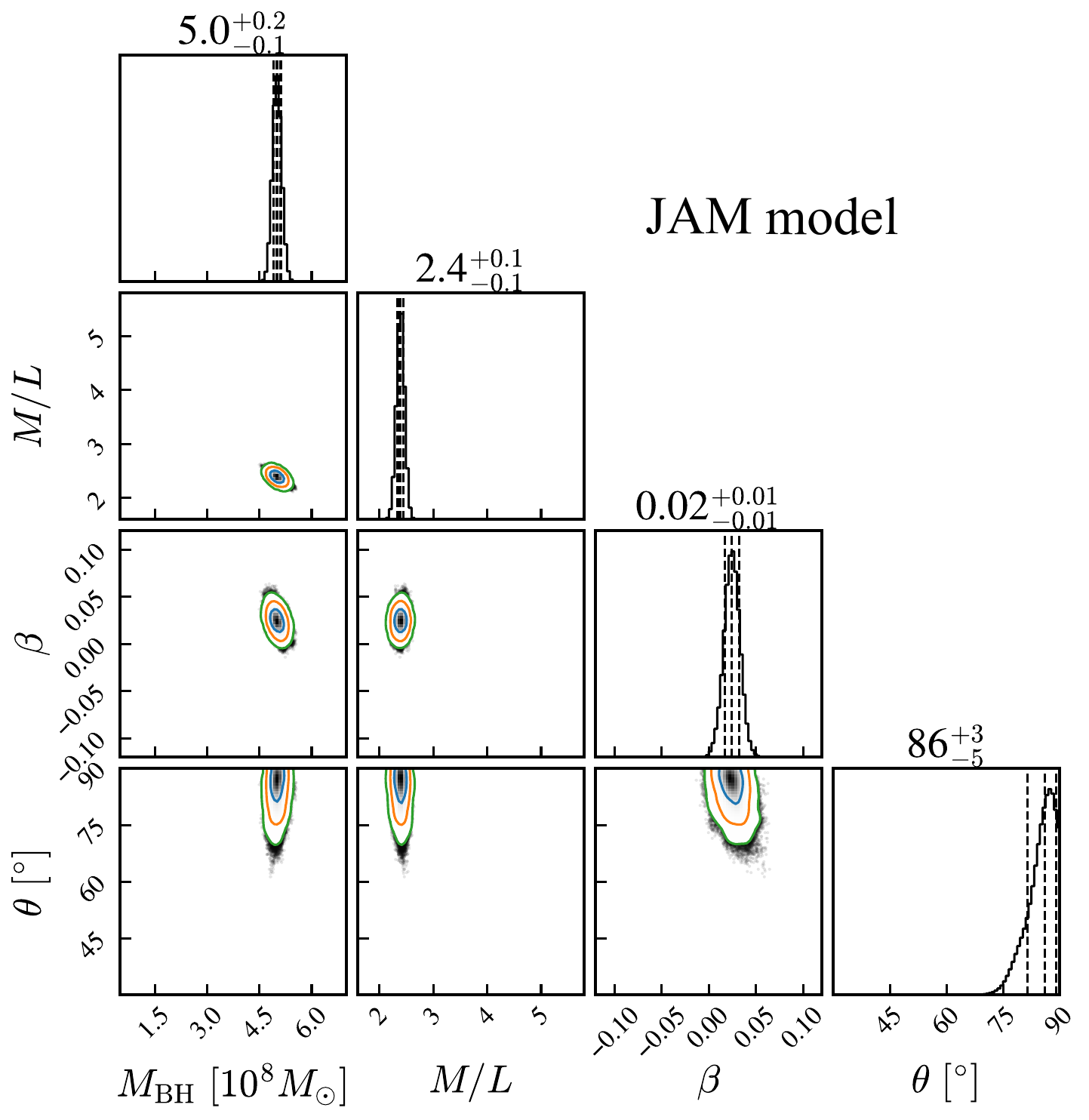}
	\caption{Posterior distributions of the model parameters derived from fits to the original data (no masking/shifting). Left: Results from the Schwarzschild orbit-superposition method, with parameters including the BH mass ($M_{\mathrm{BH}}$), stellar mass-to-light ratio ($M/L$), DM peak circular velocity ($v_{\mathrm{halo}}$), and inclination angle ($\theta$). Right: Results of the JAM model, with parameters $M_{\mathrm{BH}}$, $M/L$, velocity anisotropy ($\beta$), and $\theta$. The blue, orange, and green contours indicate the 68\%, 95\%, and 99.7\% confidence levels, respectively. Titles above the 1D histograms show the median values with their uncertainties. }%
	\label{fig:chi2}%
\end{figure*}

 \begin{figure}
	\centering	%
	\includegraphics[width=\columnwidth]{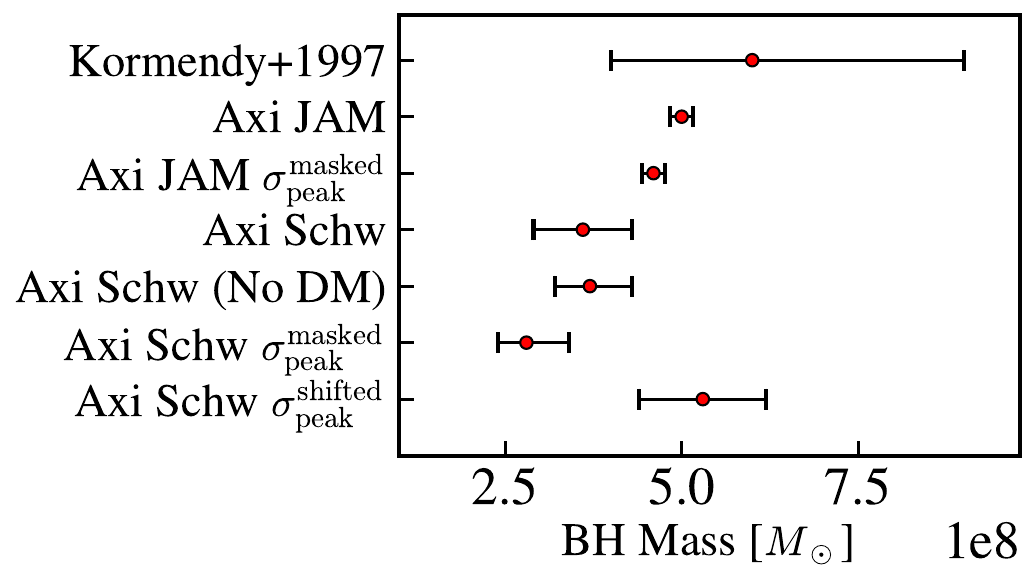}
	\caption{The BH mass measurements and their $1\sigma$ uncertainties for NGC 4486B derived from various dynamical modeling approaches. From top to bottom, the measurements correspond to: (1) the spherical isotropic Jeans modeling from \citet{Kormendy.1997}, (2) axisymmetric JAM modeling, (3) axisymmetric JAM modeling with the $\sigma$ peak masked, (4) axisymmetric Schwarzschild modeling, (5) axisymmetric Schwarzschild modeling without DM, (6) axisymmetric Schwarzschild modeling with the $\sigma$ peak masked, and (7) axisymmetric Schwarzschild modeling with the $\sigma$ peak shifted to the photometric center.}%
	\label{fig:BH_masses}%
\end{figure}

\subsection{Jeans Anisotropic Modelling}
We estimate the BH mass from the observed velocity dispersion using JAM, as implemented in the jampy Python package \citep[]{Cappellari.2008}. This method solves the Jeans equations while constructing the gravitational potential from the stellar mass distribution and a central BH, which is modeled as a Gaussian with mass \mbh. The stellar mass distribution is obtained by deprojecting the previously described MGE light profile into three dimensions, assuming a constant $M/L$. As our Schwarzschild models give only 1$\sigma$ evidence for a DM halo, we exclude a DM component in the JAM modeling and adopt a spatially constant velocity anisotropy parameter $\beta$. Under the assumption of axisymmetry, we fit four free parameters: \mbh, $M/L$, inclination angle $\theta$, and $\beta$. The modeled stellar velocity dispersion profile is obtained by integrating along the line of sight and convolving with the JWST/NIRSpec IFU PSF. We use Markov Chain Monte Carlo (MCMC) sampling to constrain all four parameters.

\section{Results and Discussion}\label{sec:result}

The second row of Fig.~\ref{fig:kinematic} shows the best-fitting Schwarzschild model based on the JWST/NIRSpec IFU kinematic dataset. The third row presents the best-fitting model for the modified dataset in which the peak of $\sigma_0$ is masked, and the fourth row illustrates the model using a kinematic map where the $\sigma_0$ peak has been manually shifted to the photometric center. The model with the shifted $\sigma_0$ shows a noticeably higher central $\sigma$ compared to the others. In contrast, the model with the masked $\sigma_0$ peak shows some differences but more closely resembles the initial model. Overall, all three models successfully recover the main features of the kinematic data, with the exception of the asymmetric features in $\sigma_0$, which cannot be captured by axisymmetric models. Note that all of these models were simultaneously fitted to the CFHT/SIS kinematics beyond $0.9\arcsec$.

Figure~\ref{fig:2set_kinem} presents the 1D kinematic profiles of $v_o$ and $\sigma_o$ from the best-fit axisymmetric model, compared with the JWST/NIRSpec IFU data and CFHT/SIS observations at larger radii ($>1$\arcsec). The model is consistent with the observed kinematics across both datasets and closely follows the CFHT/SIS measurements in the outer regions. This agreement indicates that the combined modeling approach effectively captures the galaxy's outer kinematic structure, which is essential for constraining the DM component.

Figure~\ref{fig:chi2} presents the dynamical modeling results for NGC 4486B using both the Schwarzschild and JAM approaches, illustrating the parameter constraints obtained in each case. These results correspond to fits performed on the original dataset, without any masking or shifting of data points. The figure shows marginalized one-dimensional posterior distributions along the diagonal panels and two-dimensional confidence contours for all parameter pairs, highlighting the recovered uncertainties and covariances. These distributions are constructed using likelihood-weighted kernel density estimation over the full model space.

Our JAM models yield significantly smaller formal uncertainties on BH masses compared to Schwarzschild orbit-superposition models, despite both approaches assuming axisymmetry. This discrepancy primarily stems from the stronger constraints and simplified parameterization adopted in JAM. Specifically, JAM enforces a constant anisotropy parameter, meaning it does not vary with radius, with our best-fit value of $\beta = 0.02^{+0.01}_{-0.01}$, and assumes a fixed cylindrical orientation of the velocity ellipsoid at all $(R,z)$ positions. These assumptions reduce the dimensionality of the parameter space, leading to tighter statistical confidence intervals; however, they may underestimate uncertainties by failing to capture the full range of dynamical configurations \citep[see e.g.,][]{Thater.2019, Mark.2021}. In contrast, Schwarzschild models employ large, non-parametric orbital libraries and make minimal assumptions beyond axisymmetry. They allow for radially varying anisotropy and impose no constraints on the velocity ellipsoid orientation, enabling them to explore a broader range of orbital solutions. As a result, Schwarzschild models yield more conservative but likely more realistic uncertainty estimates that better reflect the degeneracies inherent in dynamical modeling.

We measure a best-fit BH mass of \mbh$ = 3.6^{+0.7}_{-0.7} \times 10^8\,M{\odot}$. For the model based on kinematic maps with masked $\sigma_0$, we find \mbh$ = 2.8^{+0.6}_{-0.4} \times 10^8\,M{\odot}$, while the model with a shifted $\sigma_0$ yields \mbh$ = 5.3^{+0.9}_{-0.9} \times 10^8\,M{\odot}$. The results from JAM modeling are slightly higher but consistent, giving \mbh$ = 5.0^{+0.2}_{-0.1} \times 10^8\,M{\odot}$ and \mbh$ = 4.6^{+0.1}_{-0.1} \times 10^8\,M{\odot}$ for the model with masked $\sigma_0$. Figure~\ref{fig:BH_masses} summarizes and compares these dynamical BH mass measurements. The key mass component parameters and their associated uncertainties derived from the various dynamical models are summarized in Table~\ref{table:model_results}.

\begin{table*}
	\centering
	\footnotesize
	\begin{tabular}
    {l@{\hspace{20pt}}l@{\hspace{20pt}}l@{\hspace{20pt}}l@{\hspace{20pt}}l@{\hspace{20pt}}l@{\hspace{20pt}}l}
		\hline
		model    &   $ M/L \,  {\scriptstyle \mathrm{[F850LP]}}$ & \mbh\, ${\scriptstyle[10^8 \, M_{\odot}]}$  &  \mstar\,${\scriptstyle[10^9 \, M_{\odot}]}$ &  \mdm \, ${\scriptstyle[10^9 \, M_{\odot}]}$ & \mbh/\mstar & \mdm/\mstar \\  
	\hline
        JAM & $2.4^{+0.1}_{-0.1}$ & $5.0^{+0.2}_{-0.1}$ & $4.3^{+0.2}_{-0.2}$ & $0$ & $0.10-0.13$   & $0$ \\
        JAM masked $\sigma_{\mathrm{peak}}$ & $2.5^{+0.1}_{-0.1}$ & $4.6^{+0.1}_{-0.1}$ & $4.5^{+0.2}_{-0.2}$ & $0$ & $0.09-0.11$   & $0$ \\
        Schw      & $3.9^{+0.5}_{-0.4}$ & $3.6^{+0.7}_{-0.7}$ & $7.0^{+0.8}_{-0.8}$ & $1.0^{+2.0}_{-1.0}$ & $0.04-0.07$ & $\sim 0-0.5$    \\  
        Schw without DM &  $4.0^{+0.2}_{-0.3}$ & $3.7^{+0.6}_{-0.5}$ & $7.4^{+0.7}_{-0.2}$ & $0$ & $0.04-0.06$  & $0$ \\
        Schw masked $\sigma_{\mathrm{peak}}$ & $4.1^{+0.4}_{-0.5}$ & $2.8^{+0.6}_{-0.4}$ & $7.4^{+0.6}_{-0.8}$ & $1.1^{+2.6}_{-1.0}$  & $0.03-0.05$ & $\sim 0-0.6$  \\
        Schw Shifted $\sigma_{\mathrm{peak}}$ & $3.0^{+0.6}_{-0.6}$ & $5.3^{+0.9}_{-0.9}$ & $5.4^{+0.9}_{-0.8}$ & $2.7^{+9.2}_{-2.5}$ & $0.07-0.13$  & $\sim 0-2$ \\		
	\hline
	\end{tabular}
	\\
	\parbox{2\columnwidth}{\caption{Best-fit values and associated uncertainties for the dynamical mass-to-light ratio (in the F850LP band), stellar mass, BH mass, and DM mass (within 1\,$\mathrm{kpc}$), as well as the BH–to–stellar mass fraction and DM–to–stellar mass fraction, for the various dynamical models explored in this study.}
		\label{table:model_results}}
\end{table*}

All our models yield highly significant detections ($>3\sigma$) of a SMBH in NGC 4486B. The inferred BH masses range from $2.8 \times 10^8$ to $5.2 \times 10^8 \, M_\odot$, depending on the modeling approach, corresponding to $\sim 4\%$ to $13\%$ of the total stellar mass, with total stellar masses ranging from $4.3 \times 10^9$ to $7.4 \times 10^9 \, M_\odot$ across our models. Even the most conservative estimate implies that the SMBH is substantially overmassive relative to expectations from standard $M_{\mathrm{BH}}$-$M_{\star}$ scaling relations. Our dynamically derived $M/L$ in the F850LP band (approximately equivalent to $M/L_z$) range from $M/L = 2.4$ to $4.1$ across our models. For an old, metal-rich stellar population with $(z - V) \approx 1.4$\,mag (e.g., \citealt{Bruzual.2003}), this corresponds to $M/L_V \approx 8.7$–$14.9$. This is lower than the King core $M/L_V \approx 20$ reported by \citet{Kormendy.1997} for NGC\,4486B, but remains significantly above the typical $M/L_V = 4 \pm 1$ expected for old stellar populations with comparable luminosity ($M_B \sim -17$), as noted by \citet{Kormendy.1997}. This supports the interpretation that NGC\,4486B hosts an unusually dense and evolved stellar population, consistent with its compact elliptical nature. Our BH-to-stellar mass ratios ($M_{\mathrm{BH}}/M_\star \sim 4\%$--$13\%$) are also broadly consistent with their finding of $M_{\mathrm{BH}}/M_{\mathrm{bulge}} \sim 9\%$, reinforcing the conclusion that NGC~4486B hosts an unusually massive central BH.

Our Schwarzschild models do not robustly constrain the DM content within NGC 4486B, even with the inclusion of CFHT/SIS kinematic data extending to larger radii. The best-fitting Schwarzschild model based on the original dataset suggests a \mdm/\mstar$\sim 0.14$ within 1\,kpc; however, the associated uncertainties are substantial, with only an upper limit being constrained and the lower bound approaching zero (see Fig. \ref{fig:chi2}). However, including or excluding a DM component in our models does not significantly impact the estimated BH mass. The model incorporating DM yields a BH mass of \mbh$=3.6^{+0.7}_{-0.7}\,10^9 \, M_{\odot}$, while the model without DM in \mbh$=3.7^{+0.6}_{-0.5}\,10^9 \, M_{\odot}$. The slightly lower BH mass and reduced uncertainty in the model without DM are accompanied by a marginally higher stellar $M/L$, reflecting the well-known degeneracy between $M/L$ and mass content in dynamical modeling. This outcome aligns with our previous examination using mock models \citep{Behzad.2024}, indicating that in systems with an overmassive BH (\mbh/\mstar$\sim 0.1$), where the BH sphere of influence is well-resolved, the omission of a DM halo does not substantially bias the BH mass estimate (see also \citealt{Lujan.2025}).

While the axisymmetric dynamical models employed in this study allow us to constrain the BH mass in NGC 4486B, they assume equilibrium and symmetry and therefore do not fully capture the complex, non-equilibrium dynamics of the observed eccentric nuclear disk. As a result, the derived BH mass may represent a lower limit. This effect has been demonstrated in M31, where incorporating the observed asymmetries in both photometry and kinematics yielded BH masses approximately $1.5–2$ times higher \citep{Brown2013} than those derived from symmetric models \citep{Richstone1990}. 


The inclination angle is only weakly constrained in the Schwarzschild model, with a best-fit value of $72^{+11}_{-14}$ degrees. In contrast, the JAM model yields a tighter constraint of $86^{+3}_{-5}$ degrees. This difference reflects the sensitivity of axisymmetric Jeans models to the global shape of the velocity field, while the Schwarzschild model allows greater orbital flexibility. Nevertheless, the limited radial extent of the available two-dimensional kinematic data likely contributes to the overall uncertainty in both cases, reducing the model's ability to robustly constrain the viewing geometry.

In a forthcoming paper (Tahmasebzadeh et al., in preparation), we will present a more detailed investigation of the eccentric disk structure, combining high-resolution theoretical modeling and N-body simulations. This follow-up study will examine the formation, evolution, and observable signatures of eccentric disks, with the aim of developing a more comprehensive dynamical framework for interpreting double nucleus systems such as NGC 4486B.

\begin{acknowledgements}\label{sec:sum}
We are grateful to Tod Lauer for kindly providing the deconvolved F555W image from \citep{Lauer.1996}. BT and MV acknowledge funding from Space Telescope Science Institute awards: JWST-GO-02567.002-A and HST-GO-16882.002-A. EV acknowledges support from an STFC Ernest Rutherford fellowship (ST/X004066/1). M.A.T.\ and S.T.\ acknowledge the support of the Canadian Space Agency (CSA) [22JWGO1-07] and the Natural Sciences and Engineering Research Council of Canada (NSERC) [2023-03298]. MCB gratefully acknowledges support from STScI through JWST grant ERS-1364 and from the NSF through AST-2407802.

This work is based on observations made with the JWST and obtained from the Mikulski Archive for Space Telescopes (MAST). The data are available at \dataset[doi:10.17909/5wvm-0p07]{https://doi.org/10.17909/5wvm-0p07}.

\facilities{JWST(NIRSpec), HST(ACS/WFC), HST(WFPC2)}

\end{acknowledgements}

\bibliography{1.bibtex}{}
\bibliographystyle{aasjournal}



\end{document}